\documentclass[preprint]{JHEP3} 

\JHEPspecialurl{http://jhep.sissa.it/JOURNAL/JHEP3.tar.gz}
\usepackage{graphicx}
\usepackage{amsmath}

\newcommand\fverb{\setbox\fverbbox=\hbox\bgroup\verb}
\newcommand\fverbdo{\egroup\medskip\noindent%
			\fbox{\unhbox\fverbbox}\ }
\newcommand\fverbit{\egroup\item[\fbox{\unhbox\fverbbox}]}
\newbox\fverbbox

\newcommand{\ep}{\epsilon}

\newcommand{\ba}{\begin{equation}}
\newcommand{\ea}{\end{equation}}
\newcommand{\be}{\begin{eqnarray}}
\newcommand{\ee}{\end{eqnarray}}

\newcommand{\DS}{S\hspace{-4.5pt}S}
\newcommand{\DC}{C\hspace{-5.5pt}C}
\def\me{\mathcal{M}}
\newcommand{\VEC}[1]{\boldsymbol #1}
\newcommand{\lnp}[2]{\ln^{#1}(#2)}
\newcommand{\polyl}[2]{\text{Li}_{#1}(#2)}
\newcommand{\GF}[1]{\mathop{}\!\Gamma\left(#1\right)}
\newcommand{\GFP}[2]{\mathop{}\!\Gamma^#1\left(#2\right)}
\newcommand{\GENHYPGF}[5]{\mathop{}\!_{#1}F_{#2}\left[\{#3\},\{#4\};#5\right]}

\title{Integrated triple-collinear counter-terms   for the nested soft-collinear
  subtraction scheme
} 
\author{Maximilian~Delto and Kirill~Melnikov\\  Institut f{\"u}r Theoretische Teilchenphysik,
Karlsruhe Institute of Technology (KIT), Wolfgang-Gaede Stra{\ss}e 1, 76128 Karlsruhe, Germany\\
Email: \email{maximilian.delto@kit.edu, kirill.melnikov@kit.edu}
}

\received{\today} 		
\accepted{\today}		

\preprint{TTP19-004\\
P3H-19-001} 

\abstract{We obtain  analytic results for integrated triple-collinear splitting functions
  that emerge as  collinear counter-terms  in the context of the  nested
  soft-collinear subtraction scheme~\cite{Caola:2017dug}.
   With these results, all integrated subtraction terms required for NNLO QCD computations within this   scheme are 
   known analytically. In addition to improving efficiency and numerical stability of  practical computations, 
   the availability of these results  will contribute towards  establishing a general  NNLO QCD
   subtraction formula for generic hard scattering processes in hadron collisions, similar to Catani-Seymour and FKS
subtractions at NLO.}

\begin{document} 

\section{Introduction}

An  understanding  of hard processes that occur in  hadron collisions directly from first principles should allow particle
physicists to turn the Large Hadron Collider (LHC) into a precision machine and to exploit the huge dataset collected
there to
search for subtle manifestations of physics beyond the Standard Model. Perturbative QCD plays an important role in this endeavor
since it  allows for a reliable, systematically improvable description of processes with large
momentum transfer.  The development of  computational techniques for virtual loop amplitudes 
\cite{Chen:2017jvi,Gehrmann:2015ora,Caola:2014iua,Gehrmann:2011aa,Garland:2001tf,Anastasiou:2000kg}  and an improved understanding
of how to handle infra-red and collinear singularities in real emission processes
\cite{ant,njet1,njet2,grazi,Czakon:2010td,Czakon:2011ve,Herzog:2018ily,Magnea:2018hab}
resulted in a large number of theoretical
predictions for interesting hard scattering partonic  processes at the LHC, accurate  through next-to-next-to-leading
order \cite{njet1,nnloVJ,nnlottbar,nnlostop,nnloHj,nnlojj,VBF1}.
It is theoretically
interesting and phenomenologically well-motivated
to extend such computations to more complex processes; this will require significant improvements in the technology
of two-loop computations and the development of  more efficient and transparent
subtraction schemes.

The nested soft-collinear subtraction scheme introduced in Ref.~\cite{Caola:2017dug} is an attempt to
simplify and streamline the sector-improved residue subtraction scheme proposed in Refs.~\cite{Czakon:2010td,Czakon:2011ve,
 Czakon:2014oma}. The key observations made in  Ref.~\cite{Caola:2017dug} is that a significant simplification of the
subtraction terms can be achieved if subtractions are applied to gauge-invariant
scattering amplitudes rather than to individual Feynman diagrams
and if  the subtraction of soft singularities is performed first. 

Among the different subtraction terms that are required to  make  the double-real emission contribution finite, 
two genuine fully-unresolved contributions -- the double-soft  and the triple-collinear --   stand out, since
their analytic computation requires integration over  the full phase-space of two unresolved partons.
We have recently shown \cite{Caola:2018pxp}  how  the double-soft  eikonal function which is  sensitive to
momenta and color charges of all hard particles that appear in a scattering process   can be analytically
integrated over the phase-space of the unresolved  partons.

The second  double-unresolved contribution  involves triple-collinear splittings that arise when an initial
state parton emits two collinear partons before
entering the hard process or a final state parton splits into three collinear partons.   These contributions
are insensitive to the global structure of any process  and can be calculated for each of the external legs separately.
The integration of these   triple-collinear counter-terms had to be done   numerically in Ref.~\cite{Caola:2017dug}. 
Although, for practical purposes, the need to  compute the  subtraction terms numerically is hardly a serious problem, 
it is nevertheless interesting to compute these contributions  analytically.  Indeed, such a  result 
should allow to check for an analytic cancellation of the infra-red and collinear divergences in a generic hard scattering process 
and to, hopefully,  simplify the finite remainders of the integrated subtraction terms. The analytic integration of the
triple-collinear subtraction terms that appear in the context of the nested soft-collinear subtraction scheme \cite{Caola:2017dug}
is the goal of this paper. 

The remainder of the  paper is organized as follows. In Section~\ref{seq_scheme} we provide a brief overview of the
nested soft-collinear subtraction
scheme and discuss a small modification of the procedure described in Ref.~\cite{Caola:2017dug} that simplifies 
an analytic integration of the triple-collinear subtraction terms. 
In Section~\ref{seq_setup} we precisely define the triple-collinear  subtraction terms for the initial state radiation and
the phase-space over which they have to be integrated. 
In Section~\ref{seq_angint} we explain how the phase-space integrals are computed using integration-by-parts identities and
differential equations.
In Section~\ref{seq_angintSO} we discuss  how the strongly-ordered collinear contributions, required for the subtraction of sub-divergences from the triple-collinear
splitting functions,  are computed. In Section~\ref{seq_FSR}  we argue that our calculation
can be easily extended to cover the case of the collinear splittings in the final state. 
In Section~\ref{seq_res} we present the results for 
integrated triple-collinear subtraction terms.  We conclude in Section~\ref{seq_conc}.

\section{An overview of the  nested  subtraction scheme and its modification}
\label{seq_scheme}

The calculation of NNLO QCD corrections to any scattering process requires three ingredients -- two-loop virtual
corrections, one-loop virtual corrections to a process  with an additional final-state  parton  and the double-real emission contribution.
For the purpose of this paper, we are only concerned with the latter. 
We consider the double-real emission contribution to the production of a final state $X$ in the collision of two partons, i.e. 
a  partonic process $f_1+f_2 \to X + f_4 + f_5$. Here,
$X$ is the ``hard'' component of the final state, $f_{1,2}$ are the incoming partons and $f_{4,5}$ are the two
partons  that can become unresolved.  Although our discussion is applicable in general,  we will assume throughout
this paper  that  $X$ is  either a vector boson or a  Higgs boson. We will 
assume, for definiteness, that $f_{4,5}$ are gluons in this section but we will  present the results for arbitrary
integrated triple-collinear splittings later on. 

The contribution of the double-real emission
process to the cross section reads 
\be
\begin{split}
   {\rm d} \sigma_{RR}  
  &  = N \int [{\rm d} g_4] [ {\rm d} g_5] \theta(E_4-E_5) {\rm d} {\rm Lips}_X | \me (1,2;4,5;X)|^2
   \\
  &  = 
   \left \langle [{\rm d} g_4] [{\rm d} g_5 ]
   F_{\rm LM}(1,2;4,5;X) \right \rangle.
   \label{eq2.1}
   \end{split}
      \ee
      Here, $N$ is a normalization factor and ${\rm Lips}_X$ is the Lorentz-invariant phase-space
      of the final state  $X$ including the energy-momentum conserving $\delta$-function. 
The integration measure for each of the final state particles is standard except that an upper energy cut-off is introduced 
\be
   [{\rm d} g_{i}] = \frac{{\rm d}^{d-1} p_i}{(2\pi)^{d-1} 2 E_i} \theta(E_{\rm max} - E_i ),\;\;\;\; i=4,5.
   \ee
   Since such a cut-off is   not Lorentz-invariant,  Eq.(\ref{eq2.1}) should be understood 
   in the center-of-mass frame of the colliding partons.
We also note that the gluons are ordered in energy and that
   $g_4$ is always harder than $g_5$. 

The subtraction terms are  constructed iteratively, starting  from the so-called double-soft limit.
      The double-soft  limit  is defined as the limit where both $E_4$ and $E_5$ become vanishingly small but  the ratio $E_5/E_4$
   is  fixed.  We introduce the  operator  $\DS$ that extracts the double-soft limit of the differential cross section and write 
   \be
   \begin{split} 
     \langle [{\rm d} g_4] [{\rm d} g_5]  F_{\rm LM}(1,2;4,5;X) \rangle
     & = \langle
     [{\rm d} g_4] [{\rm d} g_5] ( I - \DS) F_{\rm LM}(1,2;4,5;X) \rangle
     \\
     & + \langle [{\rm d} g_4] [{\rm d} g_5] \DS F_{\rm LM}(1,2;4,5;X) \rangle.
     \label{eq2.3}
   \end{split} 
   \ee
   The first term on the right hand side does not have the double-soft singularity anymore whereas in the second term the action
   of the double-soft operator causes the 
   two gluons $g_{4,5}$ to decouple from the hard matrix element and the
   energy-momentum  $\delta$-function,  so that integration over their momenta
   can be performed. We obtain 
   \be
   \langle [{\rm d} g_4] [{\rm d} g_5] \DS F_{\rm LM}(1,2;4,5;X) \rangle 
   =    \langle F_{\rm LM}(1,2;X) \rangle \; \int [{\rm d} g_4] [ {\rm d} g_5] \; {\rm Eik}(1,2;4,5),
   \label{eq2.4}
   \ee
   where the double-soft  eikonal function is defined through the factorization formula for the matrix element squared
   \be
\lim_{E_{4,5} \to 0} |\me(1,2;4,5; X)|^2 = {\rm Eik}(1,2;4,5) \; |\me(1,2; X)|^2.
\ee
The integral of  the double-soft  eikonal function in Eq.(\ref{eq2.4}) was analytically
computed in Ref.~\cite{Caola:2018pxp} for all relevant hard emittors and unresolved partons. 

The term without the double-soft singularity in Eq.(\ref{eq2.3})
still contains the single-soft singularity that arises
in the limit $E_5 \to 0$ with $E_4$ fixed,
and various   types of collinear singularities. While the $E_5 \to 0$ singularity is treated in exactly the same
way as the double-soft  ones,  subtracting the collinear singularities requires phase-space partitioning.
The goal of this partitioning is
to ensure that only one type of collinear singularity  appears  in a given sector. This step is described
in detail in Ref.~\cite{Caola:2017dug} and we do not repeat it here.  
In this paper, we focus exclusively on the triple-collinear contributions
For example, in case
of the triple-collinear emission off the incoming parton $f_1$, such a contribution reads 
\be
\left \langle \DC_{1} \;  ( I - \sum _{k=a..d} \theta^k C_{i_kj_k} )
[{\rm d} g_4] [{\rm d} g_5 ] \; ( I - S_5) ( I- \DS) F_{\rm LM}(1,2;X;4,5) \right \rangle .
\label{eq2.6}
\ee
In the above formula, $\theta_{k \in a,b,c,d}$ stands for  partitioning functions that define the individual collinear sectors, which satisfy
\be
\sum \limits_{k=a..d} \theta_k = 1.
\ee
Moreover,  $\DC_{1}$ denotes the triple-collinear limit, where both emitted particles $g_{4,5}$  become collinear
to the incoming parton $f_1$   but there is no hierarchy of  angles between momenta $\vec p_1, \vec p_4$ and $\vec p_5$. 
In Ref.~\cite{Caola:2017dug} the triple-collinear limit was defined in the context of the parametrization introduced
in Ref.~\cite{Czakon:2011ve}. In particular, the left-most identity operator was written as a sum of  four sectors
\be
( I - \sum _{k =a..d} \theta^k C_{i_kj_k} ) \to \sum _{k =a..d} \theta^k ( I -C_{i_kj_k} ).
\label{eq2.7}
\ee
In each of these sectors  the relevant scalar products are parametrized by two variables $x_3$ and
$x_4$. The variable $x_3$ controls the relative angles between the hard emittor (say parton $1$) and the two gluons $4$ and $5$,
so that $x_3 \to 0$ limit  corresponds to the triple-collinear limit of the matrix element squared 
and the phase-space of the two gluons.\footnote{We remind the reader that the operators are supposed
  to act on anything that appears to the right of them including e.g. the phase-space of the unresolved gluons.} 
  Alternatively, we can {\it define} 
the triple-collinear operator that appears in Eq.(\ref{eq2.6}) together with Eq.(\ref{eq2.7}) as the
$x_3 \to 0$ in the parameterization that is used in each of the four sectors ~\cite{Caola:2017dug}.
On the contrary, the other parameter, $x_4$,  controls collinear  sub-divergences that arise in  different  sectors (see
Ref.~\cite{Caola:2017dug} for details).
These sub-divergences also  occur for  $x_3 \ne  0$  and are removed by $C_{i_k j_k}$ operators that appear in Eq.(\ref{eq2.6}).

The framework described above was implemented in Ref.~\cite{Caola:2017dug} where it was shown how 
Eq.(\ref{eq2.6}) can be integrated  numerically. As we already explained, the key element of the
approach described in Ref.~\cite{Caola:2017dug}
was the identification of the action of the triple-collinear operators $\DC_{1}$ and $\DC_{2}$  with the
extraction of the leading singularity that arises in the
$x_3 \to 0$ limit. This  $x_3 \to 0$ limit was taken in the matrix element, in the resolved and un-resolved
phase-spaces and in the triple-collinear splitting function.  Although
formally correct and practically
stable, this procedure impacts properties
of the unresolved phase space in a way that makes further analytic integrations very difficult.

To overcome this problem, it is useful to re-define  the triple-collinear operators $\DC_{1,2}$, making 
them  independent of the sector parametrization. Indeed, suppose we postulate that the $\DC_{1,2}$ operators
act  {\it exclusively} on the matrix element squared and on the energy-momentum  conserving
delta-function and that they  produce
 products of the triple-collinear splitting functions \cite{Catani:1999ss} and reduced matrix elements squared.
Considering collinear emissions off  parton $f_1$ for the sake of example, we  define 
\be
\DC_1 F_{LM}(1,2,4,5,X) = \frac{g_s^4}{s_{145}^2} P_{f_1 g_4 g_5}(-s_{14},-s_{15},s_{45} )
F_{LM}\left (\frac{E_1-E_4-E_5}{E_1} \cdot 1, 2; X\right).
\ee
We emphasize that in this formulation the triple-collinear operator does not act on the unresolved phase-space
and does not simplify the scalar products of four-momenta that appear as arguments of the triple-collinear
splitting function. 

It is easy to see that this reformulation leads to significant simplifications. Indeed, in this case,
the triple-collinear subtraction term is integrated  over the {\it full}  unresolved phase-space of two gluons
with additional constraint on the sum of their energies. The advantage in this procedure
is that the integrand remains a rotationally invariant function in $d-1$ spatial dimensions. We will show below that
such integrals can be mapped onto loop integrals and computed in a straightforward way 
using the method of reverse unitarity \cite{Anastasiou:2002yz}.
We also note that the strongly-ordered  subtraction terms of the form $\theta^k C_{i_kj_k}$ are computed sector by sector and rely on  a particular
choice of the phase-space parametrization. It turns out, however, that these strongly-ordered subtraction terms
simplify sufficiently to allow for a straightforward analytic integration in terms of Gamma-functions. 

Thus, a re-definition of a triple-collinear operator, compared to the original proposal in Ref.~\cite{Caola:2017dug},
allows us to complete the analytic integration of the double-unresolved
subtraction terms that arise within the framework of the
nested soft-collinear subtraction scheme. In the remaining sections of this paper, we will describe our computation
in detail and present the results for the integrated collinear counter-terms for a variety of triple-collinear splitting
functions.

\section{The setup}
\label{seq_setup}

In this section, we precisely define the triple-collinear counter-terms that need to be integrated.
As we explain below, it is convenient to use two different
energy parametrizations for  the emitted partons.   We generalize  Eq.(\ref{eq2.6}) to the case of arbitrary partons
and assume that the triple-collinear operator is defined as explained in the previous section.
We obtain 
\be
\begin{split} 
& {\cal I}_{\rm TC}  =  \bigg\langle \DC_1 \;  ( I - \sum _{k =a..d} \theta^k C_{i_kj_k} ) ( I - S_5) ( I- \DS) F_{\rm LM}(1,2;X;4,5) \bigg\rangle
\\
& = g_{s}^4\bigg\langle    ( I - \sum _{k =a..d} \theta^k C_{i_kj_k} )    ( I - S_5) ( I- \DS) 
  [{\rm d} f_4] [ {\rm d} f_5]      \frac{P_{f_1 f_4 f_5}(-s_{14},-s_{15},s_{45},z_4,z_5)}{s_{145}^2} 
 \\
 &
 \times   F_{\rm LM}\left (  \left ( 1 - \frac{E_4}{E_1}-\frac{E_5}{E_1} \right ) \cdot 1, 2 \right )  \bigg\rangle. 
 \end{split} 
\label{eq3.1}
\ee
All necessary triple-collinear splitting functions are  computed in e.g. Ref.~\cite{Catani:1999ss}.\footnote{We stress that we 
  only need spin-averaged splitting functions in the triple-collinear limits \cite{Caola:2019nzf}.
}
  We note that the minus signs
in front of some of its arguments appear because  we consider  collinear radiation
off an {\it initial} state parton.  For the same reason,
the energy fractions $z_{4,5}$ and the off-shellness of the hard parton $s_{145}$ read 
  \be
 z_{4,5} = E_{4,5}/(E_4 + E_5 - E_1),\;\;\;\; s_{145} = -s_{14} - s_{15} + s_{45}~.
  \ee

  It follows from Eq.(\ref{eq3.1}) that  the hard matrix element depends on the sum over energies of the emitted
  partons; hence, it is possible to integrate over angles of the final state partons
  $f_{4,5}$ and their  energy fractions, keeping the sum of their energies fixed.  
  It turns out to be convenient to define these energy fractions differently depending on whether or not 
  a particular final state has a double-soft  singularity.  The two different parametrizations  are discussed in
  Sections~\ref{sec:EP_set_DS}  and~\ref{sec:EP_set_no_DS}, respectively.

  \subsection{Splittings with a double-soft  singularity}
  \label{sec:EP_set_DS}
  We begin with the discussion of the energy parametrization that is mostly used for processes that exhibit  double-soft  singularities; the prominent examples include $g \to g^* + gg$ and
  $q \to q^* + gg$ initial state splittings. 
  The variable transformations  that are  used
  in this case are described in Ref.~\cite{Caola:2017dug}; for the sake of completeness we repeat this discussion  here.
  To this end, we re-write Eq.(\ref{eq3.1}) as
  \be
{\cal I}_{\rm TC} = \langle    ( I - S_5) ( I- \DS)     \;\;
 {\rm d} T_C(s_{14},s_{15},s_{45},E_4,E_5 ) \; F_{\rm LM}\left (  E_{145}/E_1 \times 1, 2 \right )  \rangle, 
\label{eq3.3}
 \ee
where $E_{145} = E_1 - E_4 - E_5$ and 
  \be
     {\rm d} T_C  =    ( I - \sum _{k =a..d} \theta_k C_k )
     [{\rm d} g_4] [ {\rm d} g_5]  \;\;    \frac{g_s^4}{s_{145}^2} \; P_{f_1 f_4 f_5 }(-s_{14},-s_{15},s_{45},z_4,z_5).
\label{eq3.4}
     \ee

     Isolating integrations over energies  we obtain 
     \be
        {\rm d} T_C = {\rm d} E_4 {\rm d} E_5 E_4^{1-2\ep} E_5^{1-2\ep} \theta(E_4 - E_5) \theta(E_{\rm max} - E_4) 
         \widetilde{T}_C(E_4,E_5,E_1),
     \ee
     where
     \be
     \widetilde{T}_C(E_4,E_5,E_1) =
     \int
( I - \sum _{k =a..d} \theta_k C_k ) \; 
    {\rm d} \Omega_{45} \;
     \frac{g_s^4}{s_{145}^2} \; P_{f_1 f_4 f_5}(-s_{14},-s_{15},s_{45},z_4,z_5),
\label{eq3.6}
 \ee
 where $\displaystyle \Omega_{45} = \frac{ {\rm d} \Omega^{(d-1)}_4 {\rm d} \Omega_{5}^{(d-1)} }{ 2^2 (2 \pi)^{2d-2}}$ is the
 angular integration measure.
 We use these expressions in  Eq.(\ref{eq3.3}) and  derive 
  \be
  \label{eq3.7}
{\cal I}_{\rm TC} =  \int \limits_{E_4 > E_5}^{}  ( I - \DS) (I-S_5)
    {\rm d} E_4 {\rm d} E_5 (E_4 E_5) ^{1-2\ep}
    \widetilde{T}_C(E_4,E_5,E_1)           F_{\rm LM}\left (  \frac{E_{145}}{E_1} \cdot 1, 2 \right )\!.
          \ee
          
          To proceed further  we need to choose an energy parametrization that decouples  hard matrix element
          from the splitting function.    In this Section, we consider
          the parametrization discussed in  Ref.~\cite{Caola:2017dug}.
For the hard collinear emission we use 
  \be
  E_{4} = E_1 (1-z)(1-r/2), \;\;\;\; E_5 = E_1 (1-z) r/2,
  \label{eq3.8}
  \ee
with $0 < z < 1$ and $0 < r < 1$. 
  This parametrization automatically satisfies the constraint $E_4 >  E_5$ and makes  hard scattering
  matrix element $r$-independent
  \be
  \langle F_{\rm LM}\left (  E_{145}/E_1 \cdot 1, 2 \right )  \rangle =
  \langle F_{\rm LM}\left (  z \cdot  1, 2 \right )  \rangle.
  \ee

  However, this parametrization is not optimal for the soft subtraction. Indeed, 
  single-soft subtraction terms that are obtained when 
  operator $S_5$ acts on  matrix elements  are computed by
  taking the limit $E_5 \to 0$ {\it at fixed} $E_4$~\cite{Caola:2017dug}.
  The parametrization in Eq.(\ref{eq3.8})  does not allow us to easily do that; for this reason, we have to switch to a
  different parametrization to describe the single-soft limit. 
  It is convenient to choose
  \be
  E_4 = E_1 (1-z), \;\;\;\; E_5 = E_1 (1-z)r.
  \ee

  After some manipulations described in Ref.~\cite{Caola:2017dug}, 
  integrated triple-collinear  subtraction terms  can  be cast into the following form 
\be
{\cal I}_{\rm TC} = [\alpha_s]^2 E_1^{-4\ep} \int \limits_{0}^{1} {\rm d} z
\left [ R_\delta \; \delta(1-z) + \frac{R_+}{\left[(1-z)^{1+4\ep}\right]_+} + R_{\rm reg}(z) \right ]
\left \langle \frac{F_{\rm LM}(z \cdot 1, 2 ) }{z} \right \rangle, \;
\ee
where $\displaystyle [\alpha_s]  = \left [\frac{\alpha_s(\mu) \mu^{2\ep} e^{\ep \gamma_E}}{2 \pi \Gamma(1-\ep)} \right ] $ is a parameter related
to the strong coupling
  constant, 
\be
\frac{1}{\left[(1-z)^{1+4\ep}\right]_+} = \left [ \frac{1}{1-z} \right ]_+ -4\ep \left [ \frac{\log(1-z)}{1-z} \right ]_+ + ...,
\ee
  and  $R_{\delta,+,\rm reg}$ read  
  \be
  \begin{split}
    & 
  R_{\delta} = \frac{ \left ( E_{\rm max}/E_1 \right )^{-4\ep} -1}{4 \ep} A_3
  - \int \limits_{0}^{1} \frac{{\rm d} r}{r^{1+2\ep}} \frac{ \left [ (1+r)^{4\ep} - 1 \right ] }{4 \ep} F(r),
  \\
  & R_+ = A_1(1) + A_2(1),
  \\
  & R_{\rm reg}(z)  = \frac{A_1(z) + A_2(z) - A_1(1) - A_2(1)}{(1-z)^{1+4\ep}}.
  \label{def_rfunc}
 \end{split} 
  \ee
  
  To write down expressions for one constant $A_1$ and three  functions $A_{2,3}$ and $F$  it is convenient to introduce
  a new operator that, acting on a function, extracts its limit when a particular variable is set to zero 
  \be
  \label{eq_limit_op}
   {\cal T}_x g(..,x,..) = \lim_{x \to 0} g(.,x,..) = g(..,0..).
  \ee
We find 
    \be
  \begin{split}
   A_1(z) & = \frac{z (1-z)^4}{2^{-2\ep}}
  \int \limits_{0}^{1} \frac{{\rm d} r}{r^{1+2\ep}} \left (1 - \frac{r}{2} \right )^{-1-2\ep}  \times 
  ( 1 - {\cal T}_r)  \Bigg  [  \left ( \frac{r}{2} \right )^2 \left ( 1 - \frac{r}{2} \right )^{2}
    \\
&   \times E_1^4 \; \widetilde{T}_C(E_1,E_1(1-z)(1-\frac{r}{2} ),E_1(1-z) \frac{r}{2} ) \Bigg ] ,
\end{split}
  \ee
  and
  \be
\begin{split} 
  &   A_2(z) = \frac{z (1-z)^4}{2\ep}
  \; \left [ 1 - \frac{\Gamma^2(1-2\ep)}{\Gamma(1-4\ep)} \right ]
  {\cal T}_r \left  [  r^2 E_1^4 \; \widetilde{T}_C(E_1,E_1(1-z),E_1(1-z)r) \right  ],
  \\
 &  A_3 = \int \limits_{0}^{1} \frac{{\rm d} r}{r^{1+2\ep}}
  \; {\cal T}_{1-z} \left [   (1-z)^4 (1 - {\cal T}_r )  \left [ r^2 E_1^4 \;  \widetilde{T}_C(E_1,E_1(1-z),E_1(1-z)r  )  \right ] \right ], \\
  &  F(r) =  {\cal T}_{1-z} \left [  (1-z)^4 r^2 E_1^4 \; \widetilde{T}_C(E_1,E_1(1-z),E_1(1-z)r  )  \right ].
\end{split} 
  \ee 
We will explain how to compute these functions in the following section. 
  
  \subsection{Triple-collinear splittings without  double-soft singularity}
  \label{sec:EP_set_no_DS}

  Several cases of  triple-collinear splittings  do not exhibit the  double-soft 
  singularities and where only one of the final state  partons may cause a  single-soft singularity. A typical
  example is  the splitting $g \to q^* + qg$.   In this case,
   we do not introduce the energy ordering  and parametrize the two energies as follows
  \be
E_{4} = E_1 (1-z)(1-r), \;\;\;\; E_5 = E_1 (1-z) r.
\label{eq3.17}
\ee
We note that the parametrization of energies in Eq.(\ref{eq3.17}) is chosen 
in such a way that $r = 0$ corresponds to the single-soft limit; this is then reflected in how $E_{4,5}$ are assigned
to final state partons.  For example,
in case of $g \to q^* + qg$ splitting,  we choose $E_5$ to be  the energy of the final state gluon and $E_4$ to be 
the energy of the final state on-shell quark.

  Following steps similar to those described in the previous section, we write the integrated triple-collinear subtraction
  term in the following way 
    \be
{\cal I}_{\rm TC} = [\alpha_s]^2 E_1^{-4\ep} \int \limits_{0}^{1} {\rm d} z \;  {\tilde R}_{\rm reg}(z) \left \langle \frac{F_{\rm LM}(z \cdot 1, 2 ) }{z} \right \rangle, 
  \ee
   where
    \be
   {\tilde R}_{\rm reg}(z)  = z (1-z)^{3-4\ep} \left[ {\tilde A}_1(z) + {\tilde A}_2(z) \right].
  \label{def_rfunc_noDS}
  \ee
  
  The two functions  that appear in Eq.(\ref{def_rfunc_noDS}) are defined as follows 
  \be
  \begin{split}
  {\tilde A}_1(z) &  =    \int \limits_{0}^{1} \frac{{\rm d} r}{r^{1+2\ep}} \left (1 - r \right )^{1-2\ep}  
  ( 1 -  {\cal T}_r)  \Bigg  [  r^2  E_1^4 \; \widetilde{T}_C(E_1,E_1(1-z)(1-r),E_1(1-z)r ) \Bigg ], \\
  {\tilde A}_2(z) &  = \frac{1}{2\ep} \left[ {  (E_{\text{max}}/E_1)^{-2\ep} } (1-z)^{2\ep} - \frac{(1-2\ep)}{(1-4\ep)}
    \frac{\GFP{2}{1-2\ep}}{\GF{1-4\ep}} \right] {\cal T}_r  \left[ r^2 E_1^4 \; \widetilde{T}_C(E_1,E_1(1-z),E_1(1-z)r) \right]. 
  \end{split}
  \ee
  Their calculation is described in the following section. 
  
   \section{Integration of  triple-collinear splitting functions}
   \label{seq_angint}

  Our goal is the analytic computation of  $R_\delta,R_+$ and $R_{\rm reg}(z)$ or $\tilde R_{\rm reg}(z)$ for all the different splitting
  functions.  There are two ingredients required for such a computation, as  can be seen 
  by inspecting Eq.(\ref{eq3.6}). One of them  is  the integral of the triple-collinear splitting
  function over the angles of unresolved gluons. The second ingredient is the integral  of the strongly-order
  collinear subtraction
  terms. These terms are particular for each of the four sectors that appear in
  Eq.(\ref{eq2.6}). Both of these contributions need to be further
  integrated over $r$, as discussed in the previous section. 
  
  We begin with the integration of the triple-collinear splitting function over the phase-space of unresolved partons.
  We consider  the following integral
  \be
    W_{abc}(E_1,E_4,E_5) =   \int {\rm d} \Omega_{45} \; P_{abc}(-s_{14},-s_{15},s_{45},z_4,z_5) / s_{145}^2 ,
    \label{eq4.1}
    \ee
    where $P_{abc}$ is a triple-collinear splitting function.  It is essential that the 
    integration  over angles in Eq.(\ref{eq4.1}) is not restricted to the collinear region; we emphasized
    this point earlier and this is an important modification
    of  the subtraction scheme described in Ref.~\cite{Caola:2017dug}.

    The resulting integral $W_{abc}$ is  a function of energies of the relevant partons. 
To compute it, we employ  methods of multi-loop computations
    such as integration-by-parts and  differential equations. The connection between
    loop and phase-space integrals is provided by reverse unitarity \cite{Anastasiou:2002yz}.
    To re-write the integral $W_{abc}$ in a way that allows one to apply reverse unitarity, we
    re-introduce integration over three-momenta of partons $4$ and $5$. We obtain
    \be
    {W}_{abc} = 
    \int [{\rm d} g_4] [{\rm d} g_5 ] \frac{\delta(k_{4}^0 - E_4) \delta( k_{5}^0 - E_5) }{ \left ( E_{4} E_{5} \right )^{1-2\ep} s_{145}^2}
    \; P_{abc}(-s_{14},-s_{15},s_{45},z_4,z_5).
    \label{eq4.2}
    \ee
    We write  partonic energies $k_{i}^0$ in the covariant form by introducing auxiliary vector $P = (1/2,\vec 0)$;
    then  $\delta(k_{i}^0 - E_i) = \delta( 2 k_i \cdot P - E_i)$.  Since $[{\rm d} g_i] = {\rm d}^dk_i \delta_+(k_i^2)$,
    we can easily turn a phase-space integral in Eq.(\ref{eq4.2}) into a loop integral by
    replacing all $\delta$-functions with the corresponding ``propagators''; we will refer to these propagators as ``cut''.
Hence, to compute  $W_{abc}$, we need to consider the following class of loop integrals
    \be
    I_{a_5,a_6,a_7,a_8}(E_1,E_4,E_5) = \int \frac{{\rm d}^dk_4}{(2\pi)^{d}} \frac{{\rm d}^dk_5}{(2\pi)^{d}}
    \frac{1}{D_1 D_2 D_3 D_4 D_5^{a_5} D_6^{a_6} D_7^{a_7}  D_8^{a_8} }, 
   \label{eq_scal_def}
    \ee
    where
    \be
    \begin{split}
      & D_1 = k_4^2,\;\;\; D_2 = k_5^2,\;\;\;D_3 = 2P\cdot k_4 - E_4,\;\;\; D_4 = 2P \cdot k_5 - E_5,
      \\
    &  D_5 = (p_1 + k_4)^2,\;\;\; D_6 = (p_1 + k_5)^2,\;\;\; D_7 = (k_4+k_5)^2, \;\;\; D_8 = \left( p_1+k_4+k_5 \right)^2.
    \end{split}
    \ee
    We note that propagators $1/D_{1,..,4}$ need to be ``cut'', in reverse unitarity sense,  so that we will only consider  integrals
    where these propagators are raised to first power.

    The importance of mapping   phase-space integrals onto two-loop
    integrals stems from the fact that we can use well-established methods to find
    algebraic relations between phase-space integrals that follow from integration-by-parts identities.
    With these relations at hand, we can identify a minimal set of integrals, the so-called master
    integrals, that have to be
    computed,  to obtain the function   $W_{abc}$.

    The master integrals
    satisfy first order differential equations in kinematic variables; solving them as an
    expansion in the dimensional regulator $\ep = (4-d)/2$ yields the required integrals up to boundary constants that
    have to be obtained using different methods.     We have used the computer program \texttt{Reduze2} \cite{vonManteuffel:2012np} to identify
    master integrals and to derive the differential equations. We note that  there exist linear relations between inverse propagators
    $D_{1,2}, D_{5,6,7,8}$ that are used to reduce the number of independent integrals. 

    For our calculation, we choose the following  set of independent master integrals
\be
 \VEC{I} = \bigg\{ I_{0,0,0,0} , I_{0,0,0,1} , I_{-1,0,0,2} , I_{0,-1,0,2}  \bigg\} .
 \label{eq_def_MI}
\ee
To derive the differential equations, it is convenient to introduce dimensionless variables
\be
E_{i} \to \omega_i = \frac{E_i}{E_1},\;\;\; i=4,5, 
\ee
and study  integrals $I_{a_4,a_5,a_6,a_7}$ as functions of $E_1$ and $\omega_{4,5}$. The dependence
of these integrals on $E_1$ follows from their mass dimensions. Hence, we define 
\be
I_{a_5,a_6,a_7,a_8}(E_1,E_4,E_5) =  E_1^{2d-6-2(a_5+a_6+a_7+a_8)} \; {\bar I}_{a_5,a_6,a_7,a_8}(\omega_4,\omega_5),
\ee
and study the dependence of the integrals $\bar I$ on $\omega_{4,5}$. 

Using integration-by-parts identities, it is straightforward to derive first-order differential equations for master integrals.
It is, however, well-known that it is beneficial to transform a system of equations to a canonical form \cite{Henn:2013pwa}.
We achieve this by applying  the algorithmic approach of Ref.~\cite{Lee:2014ioa} sequentially in both variables. 
We find 
\be
{\rm d} \VEC{{\bar J}} = \frac{\ep}{20} \sum_{i=4,5} {\rm d} \hat{M}_{\omega_i}(\omega_4,\omega_5,\ep)  \times \VEC{ {\bar J}}.
\label{eq4.8}
\ee
The linear relation between elements of the canonical basis  $\VEC{{\bar J}}$ and the original master integrals  $\VEC{{\bar I}}$
\be
\VEC{{\bar I}} = \hat{T} \VEC{ {\bar J}},
\ee
is specified by the transformation matrix 
\be
\hat T= 
\begin{pmatrix}
 \omega_4 \omega_5 & 0 & 0 & 0 \\
 0 & \frac{(1-2\ep)^2 (2 \omega_4+2 \omega_5-1)}{\ep (1-6\ep)} & -\frac{(1-2\ep)^2 (\omega_4+1)}{\ep (1-6 \ep)} & -\frac{(1-2\ep)^2 (\omega_5+1)}{\ep (1-6 \ep)} \\
 0 & -\frac{2 (1-2\ep)^2 (4 \omega_4 \ep-2 \omega_5 \ep+\ep-\omega_4)}{\ep (1-6 \ep)} & \frac{(1-2\ep)^2 (4 \omega_4 \ep-2 \ep-\omega_4)}{\ep (1-6\ep)} & -\frac{2 (1-2\ep)^2 (\omega_5+1)}{(1-6 \ep)} \\
 0 & \frac{2 (1-2\ep)^2 (2 \omega_4 \ep-4 \omega_5 \ep-\ep+\omega_5)}{\ep (1-6 \ep)} &- \frac{2(1-2\ep)^2 (\omega_4+1)}{(1-6 \ep)} & \frac{(1-2\ep)^2 (4 \omega_5 \ep-2 \ep-\omega_5)}{\ep (1-6 \ep)} 
\end{pmatrix}
\!.
\ee

The two matrices that appear in the canonical differential equation Eq.(\ref{eq4.8}) read 
\be
\begin{aligned}
\label{eq_can_dec_w4}
{\rm d} \hat{M}_{\omega_4} = &  {\rm d}\!\ln(\omega_4) \times
\begin{pmatrix} 
-40 & 0 & 0 & 0 \\
 1 & -12 & 16 & 0 \\
 -3 & 36 & -48 & 0 \\
 2 & -24 & 32 & 0 \\
\end{pmatrix}
+ {\rm d}\!\ln(\omega_4-1) \times 
\begin{pmatrix}
  0 & 0 & 0 & 0 \\
 -2 & -8 & 0 & -16 \\
 1 & 4 & 0 & 8 \\
 -4 & -16 & 0 & -32 \\
\end{pmatrix}
\\
& + {\rm d}\!\ln( \omega_4+\omega_5-1 ) \times 
\begin{pmatrix}
0 & 0 & 0 & 0 \\
 3 & -60 & 0 & 0 \\
 1 & -20 & 0 & 0 \\
 1 & -20 & 0 & 0 \\
\end{pmatrix}
+ {\rm d}\!\ln( \omega_4+\omega_5) \times 
\begin{pmatrix}
0 & 0 & 0 & 0 \\
 -2 & -24 & 16 & 16 \\
 1 & 12 & -8 & -8 \\
 1 & 12 & -8 & -8 \\
\end{pmatrix},
\end{aligned}
\\
\begin{aligned}
\label{eq_can_dec_w5}
{\rm d} \hat{M}_{\omega_5} = &    {\rm d}\!\ln(\omega_5) \times
\begin{pmatrix} 
 -40 & 0 & 0 & 0 \\
 1 & -12 & 0 & 16 \\
 2 & -24 & 0 & 32 \\
 -3 & 36 & 0 & -48 \\
\end{pmatrix}
+ {\rm d}\!\ln(\omega_5-1) \times 
\begin{pmatrix}
 0 & 0 & 0 & 0 \\
 -2 & -8 & -16 & 0 \\
 -4 & -16 & -32 & 0 \\
 1 & 4 & 8 & 0 \\
\end{pmatrix}
\\
& + {\rm d}\!\ln(\omega_4+\omega_5-1) \times 
\begin{pmatrix}
0 & 0 & 0 & 0 \\
 3 & -60 & 0 & 0 \\
 1 & -20 & 0 & 0 \\
 1 & -20 & 0 & 0 \\
\end{pmatrix}
+ {\rm d}\!\ln( \omega_4+\omega_5) \times 
\begin{pmatrix}
0 & 0 & 0 & 0 \\
 -2 & -24 & 16 & 16 \\
 1 & 12 & -8 & -8 \\
 1 & 12 & -8 & -8 \\
\end{pmatrix}.
\end{aligned}
\ee

It is straightforward to integrate this system of differential equations as an expansion in the dimensional
regularization parameter $\ep$.  However, to fully determine the integrals, we require boundary conditions.
To this end, we note that one of the master integrals -- the phase-space -- can be straightforwardly computed. It reads
\be
   {\bar I}_{0,0,0,0} = \frac{(\omega_4 \omega_5)^{1-2\ep}}{16} \left[ \Omega^{d-1}  \right]^2.
   \label{eq4.13a}
\ee
The boundary conditions for all other integrals are obtained 
by considering the limit $\omega_4 = \omega_5 = \omega \to 0$.
It follows from differential equations that the boundary conditions are  obtained from the $\omega^{n -4 \ep}$-branches of  master integrals,
where $n$ is an integer number. 
We obtain those branches by expanding denominators that appear in  master integrals in Taylor series in $\omega$. 
As an example,  we consider the  integral $I_{0,0,0,1}$; expanding the denominator in $\omega$, we obtain 
 \be
 I_{0,0,0,1} = I_{0,0,0,0} \; \omega^{-1} \;
 \int
\frac{{\rm d} \Omega_{45}  }{\left[ \Omega^{d-1} \right]^2}
 \frac{1 }{\left[\eta_{14} + \eta_{15} \right]} + ...
 \ee
 where ellipses stand for less singular terms in the $\omega$-expansion.  We compute the integral over angles and find 
\be
\begin{split}
\label{eq_boundary_int}
     \int
 \frac{{\rm d} \Omega_{45}}{\left[ \Omega^{d-1} \right]^2}
  \frac{1 }{\left[ \eta_{14} + \eta_{15} \right]} & = \frac{(1-2\ep)^2}{\ep(1-4\ep)} \frac{\Gamma^4(1-2\ep) \Gamma(1+\ep)}{\Gamma(1-4\ep) \Gamma^3(1-\ep)} \\
& - \frac{1-2\ep}{\ep}  \times \GENHYPGF{3}{2}{1,1-\ep,2\ep}{2(1-\ep),1+\ep}{-1} .
\end{split}
\label{eq4.15}
\ee
A simple analysis shows that the boundary conditions for the remaining two integrals are related to the angular integral in Eq.(\ref{eq4.15}).

The differential equations presented in Eq.(\ref{eq_can_dec_w4}) and Eq.(\ref{eq_can_dec_w5}),
together with the boundary condition  Eqs.(\ref{eq4.13a},\ref{eq_boundary_int}) provide
the starting point for the computation of  the function $\widetilde{T}_C$
for {\it any} splitting and in {\it any}  energy parameterization.
For every parameterization, we perform a change of variables in the differential equations and
solve them in terms of Goncharov Polylogarithms (GPLs) \cite{Goncharov:1994}. The resulting integrals contain GPLs of two types: 
1) $G(\{\vec a(z) \},r)$, where vectors $\vec a(z)$ is a rational function of $z$ and  2) 
$G(\{\vec b \},x)$ where $x=r,z$ and with components of $\vec b$ being rational numbers. We note that -- with
master integrals written in this way -- the integration
over the $r$-variable becomes straightforward. 
Finally, we numerically checked the master integrals by  deriving   Mellin-Barnes representations and integrating 
them  using the \texttt{MB.m} package \cite{Czakon:2005rk} for a few  values of $E_{4,5}$.

  \section{Integration of the strongly-ordered angular limits of triple-collinear splitting functions}
  \label{seq_angintSO}

  To complete the  computation of the integrated triple-collinear counter-term, we require the strongly-ordered
  subtraction terms described by the following equation
  \be
  \begin{split}
& g_s^4 \sum _{k =a..d} \left \langle     \theta_k C_k     \;\;
  [{\rm d} f_4] [ {\rm d} f_5]  \;\;    \frac{P_{f_1f_4f_5}(-s_{14},-s_{15},s_{45},z_4,z_5)}{s_{145}^2} \; 
 \; F_{\rm LM}\left ( \frac{E_{145}}{E_1} \cdot 1, 2 \right )  \right \rangle. 
 \end{split} 
\label{eq5.1}
  \ee
  This computation proceeds in the following way. For each of the four sectors, we employ the phase-space parametrization
  of  Ref.~\cite{Czakon:2010td}.  The relevant formulas for scalar products of four-momenta and for the phase-space
  in each of the four sectors can be found in the Appendix~B of Ref.~\cite{Caola:2017dug}. In terms
    of this parametrization,  the strongly-ordered angular limits correspond to $x_4 \to 0$ limit which, according
    to our earlier discussion, needs to be taken at fixed, non-vanishing $x_3$.  It is easy to check that, upon
    taking this limit, the scalar products and the unresolved phase-space simplify dramatically so that 
    integration over two remaining variables $x_3$ and $\lambda$  \cite{Czakon:2010td} can be trivially performed
    and leads to a rational function of $r$ and $z$ that can be further integrated over $r$ in a straightforward way. 
    We note that both  the integral of the triple-collinear splitting function and the integral of its strongly-ordered limit
    start at $1/\ep^2$ which, however, must cancel when the difference of the two is taken. This cancellation provides
    a welcome consistency check of the calculation.

    \section{Extension to the final state radiation}
    \label{seq_FSR}

        Up to now we have focused on the initial state radiation.  In this  section we  argue that  identical techniques can be used
        to perform analytic integration of subtraction terms for triple-collinear radiation off a final state parton. 
   For the sake of simplicity, we focus on  quark-initiated splittings 
    $q^* \to q_1 + f_4 + f_5$. In the context of the nested soft-collinear subtraction scheme, these
    final state integrated collinear counter-terms  were discussed  in Ref.~\cite{Caola:2017xuq} where the NNLO QCD corrections
    to Higgs decay to $b \bar b$ pairs were investigated.  Without going into details, we note that for the final state
    radiation the hard matrix element fully decouples.
    The integrated triple collinear limit is then a function of $\epsilon$ that is obtained by computing the following integral
        \be 
    \begin{split}
      {\cal I}_{\rm TC} = & E^{-4\ep} \bigg\{ \int \limits_{0}^{1}
      \frac{{\rm d} x_1}{x_1^{1+4\ep}} \frac{{\rm d} x_2}{x_2^{1+2\ep}}  ( 1 - {\cal T}_{x_1}) (1-{\cal T}_{x_2}) (1-x_1-x_1x_2)^{n-2\ep} \times \\
    & ~~~ \times \theta(1-x_1-x_1x_2) \left[ E^4 x_1^4 x_2^2   \widetilde{T}_C(E(1- x_1 -x_1 x_2 ), E x_1, E x_1x_2)  \right] \\
     & ~  - \frac{(E_{\text{max}}/E)^{-4\ep}-1}{-4\ep} \int \limits_{0}^{1} \frac{{\rm d} x_2}{x_2^{1+2\ep}} {\cal T}_{x_1}  (1-{\cal T}_{x_2}) \left[ E^4 x_1^4 x_2^2   \widetilde{T}_C(E(1- x_1 -x_1 x_2 ), E x_1, E x_1x_2)  \right] \bigg\},
    \end{split}
\label{eq6.1}
    \ee
    where $\widetilde{T}_C$   is obtained from a similar quantity defined in    Section~\ref{seq_setup} except that  we need to change 
    $s_{1k} \rightarrow - s_{1k} $ and $ E_1 \rightarrow - E_1$ in there. After that replacement, energies are parametrized as follows 
    \be
    E_4 = E\; x_1, ~~~  E_5 = E\; x_1 x_2 , ~~~ E_1 = E\; (1- x_1 -x_1 x_2 ).
    \ee
    In the integral Eq.~(\ref{eq6.1}), the dependence on $E_\text{max}$ is due to the cut-off theta function
        $\theta[ E_{\text{max}} - E_4 ]=\theta[ E_{\text{max}} / E - x_1 ]=\theta^E$, which is,
    however only relevant for terms that contain  the double-soft limit.
    For all other contributions, the energy conservation condition $\theta(1-x_1-x_1x_2)$ provides a
        stronger bound. To arrive at Eq.(\ref{eq6.1}),  we divide the integrand as follows
        \be
        \begin{split} 
   & \theta^E( 1 - {\cal T}_{x_1}) (1-{\cal T}_{x_2}) = \left[ (1-{\cal T}_{x_2}) -  \theta^E {\cal T}_{x_1}  (1-{\cal T}_{x_2})  \right]   \\ 
   & = \left[ (1-{\cal T}_{x_2}) - {\cal T}_{x_1}  (1-{\cal T}_{x_2})  \right] - ( \theta^E-1 ){\cal T}_{x_1}  (1-{\cal T}_{x_2}).
   \end{split}
   \ee
   We note that the second term in Eq.(\ref{eq6.1}) is generated by the $\theta^E-1$ contribution, it vanishes
   for splittings that do not exhibit a double soft singularity and in
   cases where one can choose $E_{\text{max}}=E$. This is for example the case for NNLO corrections to $1\rightarrow2$ decays.
   We also point out that a general integer power $n$ appears
   in the factor $(1-x_1-x_1x_2)^{n-2\ep}$ in Eq.(\ref{eq6.1}).  This factor  reflects possible energy-dependent damping
   factors in variants of the subtraction scheme. 
   It will be clear from the follow up discussion that
   these factors do not pose any problem for  the integration.

   To compute the angular integrals  in
   $\widetilde{T}_C$ we follow the steps outlined in Section~\ref{seq_angint}. Upon integration over angles,
    we write the result in terms of GPLs with an argument $x_2$ where appropriate; this makes the subsequent $x_2$ integration straightforward
    except that we have to respect the $\theta$-function in Eq.(\ref{eq6.1}).  We accomplish that by splitting the integral in the following
    way
    \be
    \int \limits_{0}^{1} {\rm d} x_1 \int \limits_{0}^{1} {\rm d}  x_2 \theta(1-x_1-x_1x_2)  ~~ = \int \limits_{0}^{1/2} {\rm d} x_1 \int \limits_{0}^{1} {\rm d} x_2  ~~ + \int \limits_{1/2}^{1} {\rm d} x_1 \int \limits_{0}^{(1-x_1)/x_1} {\rm d} x_2 ~.
    \ee
    The $x_2$ integration results in GPLs that contain constants and rational functions of $x_1$ both in the letters and in the arguments.
    Such GPLs can be re-written in terms of GPLs with constant letters and argument $x_1$ following 
    the procedure sometimes referred to as the "super-shuffle" \cite{Frellesvig:2018lmm}. Once this is done, the integration over $x_1$ becomes
    straightforward as well.  We note that the final result for ${\cal I}_{TC}$
    is naturally expressed in terms of GPLs of weights up to four, with rational letters and arguments; to write them
    in terms of standard set of weight-four constants, we numerically evaluate GPLs using \texttt{GINAC} \cite{Vollinga:2004sn} and use PSLQ to fit them to linear combinations of relevant transcendental and rational numbers.

    \section{Results for integrated  triple-collinear subtraction terms}
    \label{seq_res}
    
    Following  the discussion  in the previous Sections, we calculated  triple-collinear subtraction terms, for both 
    initial and final state radiation, for all partonic channels that are required to compute NNLO QCD corrections to {\it arbitrary}
    hard processes at the LHC. We illustrate these results in this Section.
    We note that all integrated triple-collinear subtraction terms  presented in this section  
    have been checked against the results of numerical integration. The complete set of results can be found   in the ancillary file provided
    with this paper.

    \TABLE[h]{%
    \centering
    \begin{tabular}{c|c|c|c}
    Splitting                        & $P_{abc}$                                                            & EP     &Name in the ancillary file     \\ \hline \hline
    $q\rightarrow ggq^*$             & $1/2\left(P_{g_4g_5q_1}+4\leftrightarrow5\right)$                    & 1      &ISR[z,1]  \\
    $g\rightarrow ggg^*$             & $1/2\left(P_{g_1g_4g_5}+4\leftrightarrow5\right)$                    & 1      &ISR[z,2]  \\
    $q\rightarrow \bar{q}'q'q^*$     & $P_{\bar{q}'_4q'_5q_1}+4\leftrightarrow5$                            & 1      &ISR[z,3]  \\
    $q\rightarrow q  q'\bar{q}'^*$   & $P_{\bar{q}'_1q'_4q_5}+4\leftrightarrow5$                            & 1      &ISR[z,4]  \\
    $q\rightarrow \bar{q}qq^*$       & $P^{\text{id}}_{\bar{q}_4q_5q_1}+1\leftrightarrow5$                  & 1      &ISR[z,5]  \\
    $q\rightarrow qq\bar{q}^*$       & $1/2\left(P^{\text{id}}_{\bar{q}_1q_4q_5}+4\leftrightarrow5\right)$  & 1      &ISR[z,6]  \\
    $g\rightarrow q\bar{q}g^*$       & $P_{g_1q_4\bar{q}_5}+4\leftrightarrow5$                              & 1      &ISR[z,7]  \\ 
    $q\rightarrow qgg^*$             & $P_{g_5q_1\bar{q}_4}$                                                & 2      &ISR[z,8]  \\
    $g\rightarrow qgq*$              & $P_{g_1g_5q_4}$                                                      & 2      &ISR[z,9]  \\
    \end{tabular}
    \label{tab:overviewISR}
    \caption{Overview of integrated triple-collinear subtraction terms for initial state splittings. In the  first column we define the splitting, in the second column, we identify the corresponding triple-collinear splitting functions in Ref.~\cite{Catani:1999ss}, where we include an additional symmetry factor where required. In the third column, we indicate which energy parametrization is used in the calculation (the energy parametrization $1$~($2$)  is described in  Sec.~\ref{sec:EP_set_DS} (\ref{sec:EP_set_no_DS}), respectively), and, finally, the  last column provides the name of the corresponding expression in the ancillary file.}
    }

    \TABLE[h]{%
    \centering
    \begin{tabular}{l|c|c|c}
    Splitting                        & $P_{abc}$                                                             &$n$ &Name in the ancillary file    \\ \hline \hline
    $q^*\rightarrow ggq$             & $1/2\left(P_{g_4g_5q_1}+4\leftrightarrow5\right)$                     &1   &FSR[1]  \\
    $q^*\rightarrow \bar{q}'q'q$     & $P_{\bar{q}'_4q'_5q_1}+4\leftrightarrow5$                             &1   &FSR[2]  \\
    $q^*\rightarrow \bar{q}qq$       & $P^{\text{id}}_{\bar{q}_4q_1q_5}+4\leftrightarrow5$                   &1   &FSR[3]  \\
    $g^*\rightarrow gq\bar{q}$       & $P_{g_1q_4\bar{q}_5}+P_{g_4q_1\bar{q}_5}+P_{g_5q_1\bar{q}_4}$         &2   &FSR[4]  \\
    $g^*\rightarrow ggg$             & $P_{g_1g_4g_5}$                                                       &2   &FSR[5]  
    \end{tabular}
    \label{tab:overviewFSR}
    \caption{Overview of integrated triple-collinear subtraction terms for final state splittings. In the  first column we define the splitting, in the second column, we identify the corresponding triple-collinear splitting functions in Ref.~\cite{Catani:1999ss}, where we include an additional symmetry factor where required. The fifth column denotes the power $n$ as defined in Eq.(\ref{eq6.1}) and the last column provides, again, the name of the corresponding expression in the ancillary file.}
    }

    As a final remark we note that  description of  the initial state radiation requires that a  crossing from final to initial states
    is performed  in the splitting functions presented  in Ref.~\cite{Catani:1999ss}. Moreover, for non-diagonal splittings $i \to j^*$
    a different color-averaging is required  when collinear limits of differential cross-sections are expressed in terms of the  splitting
    functions. To avoid potential confusion related to these subtleties, in Table~\ref{tab:overviewISR}(\ref{tab:overviewFSR})  we list the partonic channels for relevant initial(final) state splittings, the corresponding splitting functions in~\cite{Catani:1999ss} and specific details related to the energy integrations. We also emphasize that we do not introduce 
    new color averages factors into our formulas  so that the color factors in our results are the same as in the splitting
    functions in Ref.~\cite{Catani:1999ss}.

    \subsection{Initial state radiation}

    In this section we illustrate  the results of the computation by providing integrated triple-collinear subtraction terms
    for two  partonic     channels, $q \to q^* + gg$ and $g \to q^* + qg $. 
    We begin  with the presentation of  our results for the integrated triple-collinear splitting $q \to q^* + gg$ that corresponds to the
    splitting function $P_{ggq}$ in Ref.~\cite{Catani:1999ss}. Since this process exhibits  a double-soft  singularity,
    we use the parameterization of energies given  in Eq.(\ref{eq3.8}) to integrate triple-collinear subtraction terms.
    To present the result, we decompose
    it into the color factors
    \be
     R_{\delta,+,\rm reg}  = C_F^2 R_{\delta,+,\rm reg}^{\rm A}+C_F C_A \;R_{\delta,+,\rm reg}^{\rm NA},
    \ee
    and obtain
    \be
\begin{split} 
  & R_{\delta}^{\rm A} =  \frac{1}{\ep} \left ( \frac{\pi^2}{3} \ln(2) \right )   -\frac{7\pi^2}{6} \ln^2(2) + 8 \zeta_3\ln(2),
  \\
  & R_{\delta}^{\rm NA} =  \frac{1}{\ep}
  \left ( -\frac{1571}{216} + \frac{11\pi^2}{36} + \frac{3}{8} \zeta_3 + \frac{\pi^2}{3} \ln(2) + \frac{11}{2} \ln^2(2) + \left( -\frac{32}{9}+\frac{\pi ^2}{6}-\frac{11 \ln (2)}{3} \right) \ln(E_{\text{max}}/E_1) \right ) \\
  & -\frac{1}{12}\lnp{4}{2}  -\frac{176}{9}\lnp{3}{2}  -\left(\frac{79}{9}+\frac{11 \pi ^2}{12}\right) \lnp{2}{2}+\frac{513 \zeta_3+913+165 \pi ^2}{108} \ln(2)
  \\
  & + \left( \frac{64}{9}-\frac{\pi ^2}{3}+\frac{22 \ln (2)}{3} \right) \ln^2(E_{\text{max}}/E_1) \\
  & + \left( \frac{11 \zeta_3}{2}+\frac{383}{54}-\frac{22 \pi ^2}{9}-11 \ln ^2(2)+\frac{\ln (2)}{3}-\frac{2}{3} \pi ^2 \ln (2) \right) \ln(E_{\text{max}}/E_1) , \\
  & R_{+}^{A} =  -\frac{4 \pi ^2}{3} \ln(2),
  \\
  & R_{+}^{\rm NA} = \frac{1}{\ep} \left ( \frac{11 }{3} \ln(2) -\frac{\pi ^2}{6}+\frac{32}{9} \right )
  -11 \lnp{2}{2} - \frac{1+2 \pi ^2}{3} \ln(2) -7 \zeta_3 + \frac{11 \pi ^2}{9}+22.
 \end{split}   
    \ee
 Results for the regular parts are more complex. We find 
    \be
    \begin{split}
      & R_{\text{reg}}^{\rm A}  = \frac{1}{\ep} \left (
      - \frac{z+1}{2}  \ln(2) \ln(z) +  \left(1-z\right) \ln(2) +\frac{\left(z^2+3\right)}{4 (z-1)} \lnp{2}{z} -\ln(z) z+\frac{3 (z-1)}{2}
      \right )
      \\
      & +  \frac{z^2 \left(-36 \zeta_3+33+4 \pi ^2\right)-2 \left(33+2 \pi ^2\right) z-60 \zeta_3+33}{6 (z-1)} + \frac{7(z-1)}{2} \lnp{2}{2}   \\
    &  + \left(-6 z+\pi ^2 (z+1)+6\right) \ln(2) + \frac{\left(3 (z-1) z-\pi ^2 \left(3 z^2+5\right)\right)}{3 (z-1)} \ln(z)  \\
    &  + \frac{z}{2} \lnp{2}{z} + \frac{\left(9 z^2+19\right)}{12 (1-z)} \lnp{3}{z}  +\frac{7(z+1)}{4} \lnp{2}{2} \ln(z) + \frac{ \left(z^2+7\right)}{2(1-z)} \ln(2) \lnp{2}{z}  \\
    &  + (3z-1) \ln(2)\ln(z) + 6(1-z) \ln(1-z) - 4 (1-z) \ln(1-z)\ln(2)  \\
    &  +\left(-2(z+1)\ln(2)-\frac{2 \left(z^2+1\right)}{z-1} \ln(z) - 4 z\right) \; \polyl{2}{z} + \bigg( \frac{2 \left(3 z^2+5\right) }{z-1} \bigg) \; \polyl{3}{z},
    \end{split}
    \ee
    for the abelian part
    and
    \begin{align}
    & R_{\text{reg}}^{\text{NA}}  = \frac{1}{\ep} \Bigg (  \frac{\left(6 \pi ^2-61\right) z^2-15 z+76}{36 (z-1)} -\frac{11(z+1)}{6} \ln(2) + \frac{\left(11 z^2+2\right)}{12 (z-1)} \ln(z) \nonumber  \\
  & + \frac{ \left(z^2+1\right)}{2 (1-z)} \ln(1-z)\ln(z) + \bigg( \frac{1+z^2}{2(1-z)} \bigg)   \polyl{2}{z} \Bigg )
  \nonumber  \\
     & + \frac{3 \left(z^2 (48 \zeta_3 -119)-46 z-36 \zeta_3 +165\right)+\pi ^2 \left(-50 z^2+12 z+12\right)}{36 (z-1)} \nonumber \\
  &  ~ + \frac{\left(\left(61-6 \pi ^2\right) z^2+15 z-76\right)}{9 (z-1)} \ln(1-z) + \frac{\left(49 z^2+57 z-20\right)}{36 (z-1)} \ln(z)
  \nonumber \\
  &  ~ + \frac{2 \left(z^2+1\right)}{z-1} \lnp{2}{1-z} \ln(z) + \frac{(z-1)}{2} \ln(1-z)\ln(z) +\frac{\left(11 z^2+2\right)}{8(1-z)} \lnp{2}{z}
  \\
    &  ~ + \frac{2 \left(z^2+1\right)}{z-1} \ln(1-z)\ln(z)\ln(2) +\frac{22(z+1)}{3} \ln(1-z)\ln(2) + \frac{\left(z^2+1\right)}{4 (z-1)} \ln(1-z) \lnp{2}{z}  \nonumber \\
  &  ~ +\frac{11 (z+1)}{2} \lnp{2}{2} + \frac{\left(11 z^2+2\right)}{3(1-z)} \ln(2)\ln(z)
  +\frac{ \left(-7 z^2+6 z+4 \pi ^2+1\right)}{6(1-z)} \ln(2) \nonumber \\
  &  ~ + \left( \frac{2 \left(z^2+1\right)}{z-1} \ln(1-z) + \frac{2 \left(z^2+1\right)}{z-1} \ln(2) +\frac{ \left(z^2+1\right)}{2 (z-1)} \ln(z)
  +\frac{25 z^2-6 z+7}{6 (z-1)}\right) \; \polyl{2}{z} \nonumber \\
  &  ~ + \bigg(\frac{2 \left(z^2+1\right)}{z-1}\bigg) \;\polyl{3}{1-z} + \bigg( \frac{\left(z^2+1\right)}{2(1- z)} \bigg) \; \polyl{3}{z},
  \nonumber 
\end{align}
for the non-abelian. As the second example, we present the results for the integrated triple-collinear counter-term that describes the $g \to q^* + qg$ splitting.
Since in this case  there is no double-soft  singularity, we use the energy parametrization given in Eq.(\ref{eq3.17}). Another consequence
of the absence of the double-soft singularity is the regularity of the counter-term so that it  does not contain either
a $\delta(1-z)$ function  or a  plus-distribution.  We decompose  the result into the color factors 
     \be
     {\tilde R}_{\text{reg}} =  C_F^2  \tilde{R}_{\text{reg}}^{\text{A}}+C_F C_A  \tilde{R}_{\text{reg}}^{\rm NA} 
     \ee
and find 
    \begin{align}
    & {\tilde R}_{\text{reg}}^{\text{A}}   = \frac{1}{\ep} \Bigg (  \frac{8 \pi ^2 z^2-8 \pi ^2 z-15 z+4 \pi ^2-3}{12} + 3 \left(2 z^2-2 z+1\right) \ln(1-z)\ln(2) \nonumber \\
    & + \left(-2 z^2+2 z-1\right) \ln(1-z) \ln(z)  +\frac{1-2 z}{2} \ln(z) \ln(2)  +\frac{-9 z^2+11 z-5}{2} \ln(2)   \nonumber \\
    &+\frac{4 z^2-6 z+3}{4} \lnp{2}{z}-\frac{3 }{4} \ln(z) - \bigg(2 z^2-2 z+1 \bigg) \polyl{2}{z} \nonumber \\
    &-3 (1-2z+2z^2) \ln(2) \ln(E_{\text{max}}/E_1)  \Bigg ) \nonumber \\
    & + \frac{-3 \pi ^2 z^2+12 z \zeta_3+3 \pi ^2 z-24 z-6 \zeta_3-\pi ^2}{3} -9  \left(2 z^2-2 z+1\right) \lnp{2}{1-z} \ln(2) \nonumber \\
    & +4  \left(2 z^2-2 z+1\right) \lnp{2}{1-z} \ln(z) -\frac{19\left(2 z^2-2 z+1\right)}{2} \ln(1-z) \lnp{2}{2}  \nonumber \\
    & +4 \left(2 z^2-2 z+1\right) \ln(1-z)\ln(2) \ln(2) +\left(18 z^2-22 z+7\right) \ln(1-z)\ln(2) \\
    & +\frac{\left(2 z^2-2 z+1\right)}{2} \ln(1-z) \lnp{2}{z} + \ln(1-z)\ln(z) +\frac{7(2 z-1)}{4} \ln(z)\lnp{2}{2} \nonumber \\
    & +\frac{3-4 \pi ^2 z^2+4 \pi ^2 z+15 z-2 \pi ^2}{3} \ln(1-z) + \frac{57 z^2-71 z+32}{4} \lnp{2}{2}  \nonumber \\
    & +\frac{-8 z^2+14 z-7}{2} \lnp{2}{z} \ln(2) + 2 (z+2) \ln(z) \ln(2) \nonumber \\
    & +\frac{-4 \pi ^2 z^2-117 z^2+8 \pi ^2 z+150 z-4 \pi ^2-27}{6} \ln(2) +\frac{-28 z^2+38 z-19}{12} \lnp{3}{z}\nonumber \\
    &+\frac{(8 z+9)}{8} \lnp{2}{z} + \frac{-32 \pi ^2 z^2+40 \pi ^2 z-21 z-20 \pi  ^2+9}{12} \ln(z)   \nonumber \\
    &+  \bigg(\ln(2) \left(8 z^2-12 z+6\right)+\left(-2 z^2+2 z-1\right) ( \ln(z) - 4 \ln (1-z) ) -2\bigg) \polyl{2}{z} \nonumber \\
    & +\bigg(8 z^2-8 z+4\bigg) \polyl{3}{1-z} + \bigg(14 z^2-18 z+9\bigg)  \polyl{3}{z} \nonumber  \\
    &+3 (1-2z+2z^2) \ln(2) \ln^2(E_{\text{max}}/E_1) \nonumber \\
    &+ \bigg( \frac{19(1-2z+2z^2)}{2} \ln^2(2) + 6 (1-2z+2z^2) \ln(1-z)\ln(2) + 3 \ln(2) \nonumber \\
    & - \frac{2\pi^2 (1-2z+2z^2) }{3} \bigg) \ln(E_{\text{max}}/E_1) \  ,
    \end{align}
and
    \begin{align}
      & {\tilde R}_{\text{reg}}^{\text{NA}} = \frac{1}{\ep} \Bigg ( \frac{-6 \pi ^2 z^3-67 z^3+3 \pi ^2 z^2+81 z^2-3 \pi ^2 z-27 z+13}{9 z}
      \nonumber \\
      & ~+\left(2 z^2-2 z+1\right) \ln(1-z) \ln(2) 
      +\left(2 z^2-2 z+1\right) \ln(1-z) \ln(z) \nonumber \\
      &  - \left(2 z^2+2 z+1\right) \ln(1+z)\ln(z) + (4 z+1) \ln(z)\ln(2)
       +\frac{4-31 z^3+24 z^2+3 z}{6z} \ln(2) \nonumber \\
       &  ~ +\frac{6 z+1}{2} \lnp{2}{z}+\frac{ 12 z+1}{2} \ln(z)
        - \bigg(2 z^2+2 z+1\bigg) \polyl{2}{-z}+\bigg(2 z^2-2 z+1\bigg) \;\polyl{2}{z} \nonumber \\
        & - (1-2z+2z^2) \ln(2) \ln(E_{\text{max}}/E_1)  \Bigg ) \nonumber \\
    &  +\bigg( \left(8 z^2+8 z+4\right) \big( \ln(1-z) + \ln(2) \big)  + \left(2 z^2-6 z+1\right) \ln(z) \bigg) \polyl{2}{-z} \nonumber \\
    & + \bigg(\left(-8 z^2+8 z-4\right)\ln(1-z)-8 (z-3) z \ln(2) -4 z\ln(z) \bigg )  \polyl{2}{z}  \\
    &   +\frac{44 z^3+48 z^2+15 z+8}{3 z} \; \polyl{2}{-z} + \frac{-22 z^3+96 z^2-3 z+20}{3 z}  \polyl{2}{z} \nonumber \\
    & -\bigg(18 z^2-2 z+9\bigg)  \polyl{3}{1-z} + \bigg(10 z^2+26 z+5\bigg)  \polyl{3}{-z} \nonumber \\
    & + \bigg(4 z^2+4 z+2 \bigg) \left (  3 {\rm Li}_3 \left ( \frac{z}{1+z} \right ) + \polyl{3}{1-z^2} \right ) +\bigg(32 z+4\bigg)  \polyl{3}{z} \nonumber \\
    & +  (1-2z+2z^2) \ln(2) \ln^2(E_{\text{max}}/E_1) \nonumber \\
    & + \bigg( \frac{7(1-2z+2z^2)}{2} \ln(2) + 2 (1-2z+2z^2) \ln(1-z) +  1  \bigg) \ln(2) \ln(E_{\text{max}}/E_1)  .
 \nonumber 
    \end{align}
    Results for other integrated triple-collinear counter-terms are of a similar complexity.
    They can be found in an ancillary file provided with this paper.
    
    \subsection{Final state radiation}

    In this section we present the results for the integrated triple-collinear counter-terms
    relevant for two partonic channels,  $q^* \rightarrow ggq$ and $q^* \rightarrow \bar{q}q' \bar q'$.
    Following    Ref.~\cite{Catani:1999ss}, we split the three-quark triple-collinear splitting function into two contributions 
    \be
    P_{\bar{q}_1q_2q_3} = P_{\bar{q}'_1q'_2q_3} + P^{\text{id}}_{\bar{q}_1q_2q_3},
    \ee
    that allow a description of final states with both identical and different quark flavors. We present the corresponding contributions
    separately. 

    To present the results, we define
    \be 
    {\cal I}_{\rm TC}^{qab} =  [\alpha_s]^2 E^{-4\ep} R^{qab},
    \ee
    and obtain
\begin{align}
  R^{qgg} &=C_A C_F \;\Bigg\{   \frac1{\ep} \bigg[ -\frac{1015}{108} + \frac{19 \zeta_3}{8}+\frac{\pi ^2}{8}+\frac{11 }{2}\ln ^2(2)
    -\frac{11 }{4} \ln (2) +\frac{1}{3} \pi ^2 \ln (2) \bigg] \nonumber \\
   & ~~~~ + \bigg[ -\frac{2281}{48} -2 \polyl{4}{1/2}+\frac{25 \zeta_3}{24}-\frac{13}{4} \zeta_3 \ln (2)-\frac{119 \pi ^2}{144}+\frac{173 \pi ^4}{480}-\frac{\ln ^4(2)}{12} \nonumber \\
      & ~~~~~~~ -\frac{176 }{9} \ln ^3(2) -\frac{19 }{36} \ln^2(2) -\frac{11}{12} \pi ^2 \ln^2(2)-\frac{1247 }{108} \ln(2)
      +\frac{161}{36} \pi ^2 \ln (2) \bigg] \Bigg\}   \\
    & + C_F^2 \;\Bigg\{ \frac1{\ep} \bigg[ \frac{31}{16} -2 \zeta_3+\frac{9}{8} \ln (2)+\frac{1}{3} \pi ^2 \ln (2) \bigg]
    + \bigg[ \frac{715}{32} + 16 \zeta_3 \ln (2)-\frac{7 \pi ^4}{30} \nonumber \\
 & ~~~~~~~  -\frac{63}{16} \ln^2(2) -\frac{7}{6} \pi ^2 \ln^2(2) + \frac{17 }{8} \ln(2) +\pi^2 \ln (2) \bigg]  \Bigg\},  \nonumber \\
    R^{q\bar{q}'q'}&=  C_F T_R \;\Bigg\{ \frac1{\ep} \bigg[ \frac{329}{108}-2 \ln^2(2)+\ln (2) \bigg] + \bigg[ \frac{2773}{216} + \frac{19 \zeta_3}{6}  \\
      & + \frac{35 \pi ^2}{72}+\frac{64}{9}  \ln ^3(2)  +\frac{32 }{9}\ln ^2(2)+\frac{43 }{27}\ln (2)-\frac{13}{9} \pi ^2 \ln (2)  \bigg] \Bigg\} ~, \nonumber \\
    R^{q\bar{q}q,\text{id}}&=C_F\left( C_F - \frac{1}{2}~C_A \right) \;\Bigg\{ \frac1{\ep} \bigg[ -\frac{13}{4}-2 \zeta_3+\frac{\pi ^2}{2} \bigg]
     \\
     & + \bigg[ -\frac{335}{8} + 39 \zeta_3+8 \zeta_3 \ln (2)
        +\frac{5 \pi ^2}{3} -\frac{14 \pi ^4}{45}+13 \ln (2)-2\pi ^2 \ln (2) \bigg] \Bigg\} ~. \nonumber 
\end{align}

    \section{Conclusion}
    
    \label{seq_conc}

    In this paper, we computed all the relevant integrated  
    triple-collinear  subtraction terms for both initial and final state radiation 
    in the context of the
    nested soft-collinear subtraction scheme of Ref.~\cite{Caola:2017dug}.  
    Together with  the results for the integrated double-soft  eikonal factor presented earlier in Ref.~\cite{Caola:2018pxp},
    the computation reported in  this paper
    completes the calculation  of the required
    integrated subtraction terms for the subtraction scheme of Ref.~\cite{Caola:2017dug}. In addition to improving the efficiency and
    numerical stability of  practical  computations, these results should enable the derivation of a general NNLO QCD subtraction formula for
    arbitrary hard processes at the LHC, similar to Catani-Seymour \cite{Catani:1996vz}
    and FKS \cite{Frixione:1995ms,Frixione:1997np} schemes at NLO QCD.

\section*{Acknowledgments} 
We are grateful to F.~Caola and R.~Roentsch for useful conversations. We would like to thank Ch.~Wever for his help with  Mellin-Barnes
integration.  The research of K.M. is supported by BMBF grant 05H18VKCC1 and by the DFG Collaborative Research Center TRR~257 ``Particle Physics
Phenomenology after the Higgs Discovery''. M.D. is supported by the start-up funds provided by Karlsruhe Institute of Technology.

\end{document}